\documentclass[pra,onecolumn,floatfix,superscriptaddress,longbibliography,notitlepage, nofootinbib]{revtex4-1}
\pdfoutput=1

\usepackage[utf8]{inputenc}
\usepackage{braket}
\usepackage{amsmath}

\usepackage{dcolumn}   % needed for some tables
\usepackage{bm}        % for math
\usepackage{amssymb}
\usepackage{mathtools}
\usepackage{tikz}
\usepackage{qcircuit}
\usepackage{appendix}
\usepackage{hyperref}
\usepackage{subcaption}
\usepackage{amsmath,amsfonts,amssymb}
\usepackage{mathtools}
\usepackage{thmtools,thm-restate}
\usepackage{algorithm}
\usepackage{mathrsfs}
\usepackage{tikz}
\usepackage{subcaption}
\usepackage{pgfplots}
\usetikzlibrary{3d}
\usepackage{tkz-euclide}

\usepackage{algpseudocode}

\newcommand{\Ibb}{\mathbb{I}}

\newtheorem{lemma}{Lemma}

\usetikzlibrary{arrows.meta}

\def\BibTeX{{\rm B\kern-.05em{\sc i\kern-.025em b}\kern-.08em
    T\kern-.1667em\lower.7ex\hbox{E}\kern-.125emX}}

\begin{document}
\title{(Nearly) Optimal Time-dependent Hamiltonian Simulation}
\author{Nhat A. Nghiem}
\email{nhatanh.nghiemvu@stonybrook.edu}
\affiliation{Department of Physics and Astronomy, State University of New York at Stony Brook, Stony Brook, NY 11794-3800, USA}
\affiliation{C. N. Yang Institute for Theoretical Physics, State University of New York at Stony Brook, Stony Brook, NY 11794-3840, USA}

\begin{abstract}
We describe a simple quantum algorithm to simulate time-dependent Hamiltonian, extending the methodology of quantum signal processing. The framework achieves optimal scaling up to some factor with respect to other parameters, and nearly optimal in inverse of error tolerance, which could be improved to optimal scaling under certain input models. As applications, we discuss the problem of simulating generalized lattice system and time-periodic, or Floquet system, showing that our framework provides a neater yet highly efficient solution, achieving optimal/nearly optimal scaling in all parameters. In particular, our method also paves a new way for studying phase transition on quantum computer, extending the reach of quantum simulation. 
\end{abstract}
\maketitle

\noindent
\textbf{Introduction.} Quantum computation was largely motivated by the quest to simulate nature. As Feynman remarked in \cite{feynman2018simulating}, nature is inherently quantum, and thus, its simulation should also be driven by quantum mechanisms. This insight has spurred significant advancements in quantum computation and information, with quantum simulation emerging as one of its major applications. The foundational works of \cite{lloyd1996universal} and \cite{aharonov2003adiabatic} demonstrated how to simulate time-independent Hamiltonians composed of local interactions. Building on these results, numerous studies have refined and expanded simulation techniques to handle various quantum systems under different input models, incorporating methods such as sparse access \cite{berry2007efficient}, quantum walk \cite{berry2015hamiltonian, berry2015simulating, childs2010relationship}, and the linear combination of unitaries \cite{childs2012hamiltonian}. Beyond the time-independent setting, several works have explored simulations in the time-dependent regime \cite{poulin2011quantum, berry2020time, chen2021quantum, an2022time, low2018hamiltonian, kieferova2019simulating}.

Feynman’s insight highlights the importance of leveraging physical intuition in designing quantum algorithms. This philosophy played a crucial role in the development of quantum signal processing, a technique from quantum control that enables Hamiltonian simulation with optimal complexity, as demonstrated in the seminal work of \cite{low2017optimal}. Here, complexity is measured by the number of elementary operations, such as single-qubit gates and queries to the Hamiltonian under consideration. Building on this success, a higher-dimensional generalization known as qubitization \cite{low2019hamiltonian} was introduced, further optimizing Hamiltonian simulation. Beyond quantum simulation, quantum signal processing and qubitization have given rise to a powerful framework known as quantum singular value transformation (QSVT) \cite{gilyen2019quantum}, which unifies a broad class of quantum algorithms.

Within the context of quantum simulation, Ref. \cite{low2017optimal} primarily addressed time-independent Hamiltonians. However, time-dependent Hamiltonians encompass a broader class of systems, and it is reasonable to expect that efficient simulation techniques should also be guided by physical intuition. At first glance, the approach in \cite{low2017optimal} appears unsuitable for the time-dependent setting, as the Hamiltonian's spectrum is no longer static. Nevertheless, we show that it is possible to extend the principles of quantum signal processing—based on single-qubit operations as computational modules—to the time-dependent case. In particular, our approach achieves optimal complexity with respect to the most relevant parameters, thereby inheriting the simplicity and efficiency of quantum signal processing for time-independent simulations.

\noindent
\textbf{Main Framework.} The dynamics of a quantum system under time-dependent Hamiltonian is governed by Schrodinger's equation $\frac{\partial \ket{\psi} }{\partial t} = -i H(t) \ket{\psi}$, where $H(t)$ is the (time-dependent) Hamiltonian. In an equivalent Heisenberg picture, the time-evolution operator $U(t)$ is defined by $\frac{\partial U(t)}{\partial t} = -i H(t) U(t)$. The problem of simulating Hamiltonian is to use a universal quantum computer to construct the unitary operator $U$ such that:
\begin{align}
    || U - \tau \exp(-i \int_0^t H(s) ds) || \leq \epsilon
\end{align}
where $||.||$ refers to the spectral norm of matrix and $\tau$ is the time-ordering operator. In the time-independent case, the integral $ \exp(-i \int_0^t H(s) ds) $ becomes $\exp(-i H t)$. Without loss of generality, we consider the simulation at time $t$ for $0 \leq t \leq 1$. We consider the case where the $N$-qubit time-dependent Hamiltonian is decomposed as $H(t) = \sum_{i=1}^m \gamma_i(t) H_i$, where $m$ is some known factor, each $H_i$ is an $N$-qubit time-independent Hamiltonian with $||H_i|| \leq 1/2$, and $\gamma_i(t)$ is a time-dependent coefficient that is a (computable) function of $t$ with norm $|\gamma_i(t)| \leq 1$ for $0 \leq t \leq 1$. Additionally, let the Hamiltonian $H(t)$ be commuted at different times, i.e. $H(t_1) \neq H(t_2)$ for $t_1 \neq t_2$. Suppose further that each $H_i$ can be (efficiently) simulated, for example, by by sparse-access method \cite{berry2007efficient}, truncated Taylor series \cite{berry2015simulating}, quantum walk \cite{berry2015hamiltonian}, etc. It means that our framework (soon to be outlined) is applicable whenever we can efficiently simulate the Hamiltonian under decomposition. To proceed, we have the following:
\begin{align}
    \exp(-i \int_0^t H(s) ds) %&= \exp( -i \int_0^t (  \sum_{i=1}^m \gamma_i(s) H_i ) ds )  \\
    &= \exp( -i (  \sum_{i=1}^m \int_0^t  \gamma_i(s)ds ) H_i ) 
\end{align} 
Given that $\gamma_i(t)$ is a known function of $t$, then its integral is efficiently analytical integrable. Denote $\int_0^t \gamma_i(s)ds = \alpha_i(t)$. 

A basic operation in quantum information processing is single qubit rotation. For example, a rotation around $x$ axis has the matrix form: 
\begin{align}
    W(x) = \exp( i \sigma_x \arccos(x) ) = \begin{pmatrix}
        x & i \sqrt{1-x^2} \\
        i \sqrt{1-x^2} & x 
    \end{pmatrix}
\end{align}
Quantum signal processing \cite{low2017optimal} is built upon the use of $W(x)$ and another simple single qubit rotation along $z$ axis to generate a matrix whose entries are polynomial of $x$. More concretely, we have the following:
\begin{align}
    W_{\Phi }(x) &= \exp(-i\theta_0 \sigma_z) \prod_{i=1}^k W(x) \exp(i\theta_1 \sigma_z)  \\
                &= \begin{pmatrix}
                    P(x) &  iQ(x) \sqrt{1-x^2} \\
                    iQ^*(x) \sqrt{1-x^2} & P^*(x) 
                \end{pmatrix}
                \label{eqn: W}
\end{align}
where $P(x)$ and $Q(x)$ are polynomial of degree at most $k$ and $k-1$, respectively. A key property of the above transformation is that, given degree and coefficients of $P(x)$ (and $Q(x)$), then the rotational angles $\Phi = ( \theta_0,\theta_1, ...,\theta_k  )$ can be computed algorithmically \cite{gilyen2019quantum}. We remark an important notation from \cite{gilyen2019quantum} that, in the above operator, if we only pay attention to the top left corner and ignore the remaining part, then $W_{\Phi}$ is said to be a block encoding of $P(x)$. A more precise definition of block encoding is as follows. Suppose that $A$ is an s-qubit operator, $\alpha, \epsilon \in \mathbb{R}_+$ and $a \in \mathbb{N}$, then we say that the $(s+a)$-qubit unitary $U$ is an $(\alpha, a ,\epsilon)$-block encoding of $A$, if
    $$ || A - \alpha (\bra{0}^{\otimes a} \otimes \mathbb{I}) U ( \ket{0}^{\otimes a} \otimes \mathbb{I} ) || \leq \epsilon$$
An important algebraic property of block encoding, as detailed in \cite{gilyen2019quantum} (Lemma 52 and Lemma 53), is that given multiple block encoding of operators, one can form block encoding of arbitrary linear superposition, and product of these operators. 

At a given time $t$, a single qubit rotation with angle $\arccos(t)$ has matrix form:
\begin{align}
    W(t) = \begin{pmatrix}
        t & i \sqrt{1-t^2} \\
        i \sqrt{1-t^2} & t 
    \end{pmatrix}
\end{align}
Suppose that for each $\alpha_i(t)$, there is a polynomial approximation of $\alpha_i(t)$ and that polynomial is constructable. As described in \cite{gilyen2019quantum}, a wide range of functions could be efficiently approximated by some polynomial, in a sense that the degree of polynomial is logarithmical in the error desired. Then the following equation \ref{eqn: W}, we can determine the angles $\Phi^i$ (the supersript $i$ is to specifically denote the angles corresponding to polynomial approximation of $\alpha_i(t)$) such that:
\begin{align}
\label{eqn: 7}
    W_{\Phi^i}(t) = \begin{pmatrix}
        P_i(t)  & \cdot \\
        \cdot & \cdot
    \end{pmatrix} = \begin{pmatrix}
        \alpha_i(t) & \cdot \\
        \cdot & \cdot
    \end{pmatrix}
\end{align}
where we use $(.)$ to emphasize that it is irrelevant part, as we only care about the top left corner $\alpha_i(t)$. The above matrix is defined as block encoding of $\alpha_i(t)$. To proceed, given that the simulation of $H_i$ can be achieved as we have assumed, we use the following result of \cite{gilyen2019quantum}:
\begin{lemma}
\label{lemma: logarithmicofunitary}
    Suppose that $U = exp(-iH)$, where $H$ is a Hamiltonian of (spectral) norm $||H|| \leq 1/2$. Let $\epsilon \in (0,1/2]$ then we can implement a $(2/\pi, 2,\epsilon)$-block encoding of $H$ with $\mathcal{O}( \log(1/\epsilon)$ uses of controlled-U and its inverse, using $\mathcal{O}(\log(1/\epsilon)$ two-qubits gates and using a single ancilla qubit. 
\end{lemma}
to construct the block encoding of $\pi H_i/2$. Hence, the tool developed in \cite{camps2020approximate} allows us to construct the block encoding of $\pi P_i(t) H_i/2$, given that we have described earlier how to obtain the block encoding of $\pi H_i/2$. Then we can use lemma 52 in \cite{gilyen2019quantum} to construct the block encoding of $\frac{\pi}{2m} \sum_{i=1}^m \alpha_i(t) H_i$. \\

From such block encoding, we can use again the result of \cite{gilyen2019quantum, low2017optimal} to construct the block encoding of $\exp\big(- i \frac{\pi}{2m} \sum_{i=1}^m \alpha_i(t) H_i\big) $, making use of the so-called Jacobi-Anger expansion. The number of usage of block encoding of $ \sum_{i=1}^m \frac{1}{m} \pi \alpha_i H_i $, as analyzed in \cite{low2017optimal}, is $\mathcal{O}\Big( \frac{\log(1/\epsilon)}{ \log( e + \log(1/\epsilon) )}  \Big)$. Then by using $\mathcal{O}(m)$ block encoding of $\exp\big(-  i \frac{\pi}{2m} \sum_{i=1}^m \alpha_i(t) H_i \big)$, combining with lemma 53 in \cite{gilyen2019quantum} allows us to construct the block encoding of their products, which is our desired operator $\exp\Big(-i \sum_{i=1}^m \alpha_i(t) H_i \Big)$. 

To analyze the complexity, we remind that there are four key steps in our framework. First, one choose certain angle $\arccos(t)$ for single qubit rotation gate to obtain the operator
\begin{align}
    W(t) = \begin{pmatrix}
        t & i \sqrt{1-t^2} \\
        i \sqrt{1-t^2} & t 
    \end{pmatrix}
\end{align}
Then one uses quantum signal processing to transform the above operator into: 
\begin{align}
   \begin{pmatrix}
        t & i \sqrt{1-t^2} \\
        i \sqrt{1-t^2} & t 
    \end{pmatrix} \longrightarrow \begin{pmatrix}
        \alpha_i(t)  & \cdot \\
        \cdot & \cdot
    \end{pmatrix}
\end{align}
Let $T_W$ be the time required to implement a single rotation gate $W(t)$. The complexity of this step, which is the number of times $W(t)$ being used plus the time required to implement $W(t)$, is $\mathcal{O} \big( T_W \deg(\alpha_i (t)) \big)$ (we have ignored the error dependence as we assumed that $\alpha_i$ is efficiently realized with modest error). The next step is to obtain the block encoding of $\pi H_i/2$ for all $i$, which is possible based on the (assumed) ability to obtain the exponent $\exp(-i H_i)$, or more precisely $\exp(-i \pi H_i/2 )$ in this case. Since $H_i$ is time-independent, standard techniques can be applied, and optimal complexity has been found \cite{berry2015hamiltonian, low2017optimal}, which is $\mathcal{O}(  d_i ||H_i|| + \log \frac{1}{\epsilon} )$ where $d_i$ is the sparsity of $H_i$. From lemma \ref{lemma: logarithmicofunitary}, the complexity of obtaining block encoding of $\pi H_i/2$ is $\mathcal{O}( ( d_i ||H_i|| + \log \frac{1}{\epsilon} ) \log \frac{1}{\epsilon} )$. The next step is to form the block encoding of linear combination $\sum_{i=1}^m \frac{\pi}{2m} \alpha_i H_i $. The construction of block encoding of $\alpha_i(t) \pi H_i/2$ takes a single use of block encoding of $\alpha_i(t)$ and $\pi H_i/2$, hence the complexity of obtaining it is $\mathcal{O}\Big( T_W  \big(\deg \alpha_i(t) \big) \big( d_i ||H_i|| + \log \frac{1}{\epsilon} \big) \log \frac{1}{\epsilon}  \Big)$. It takes a single use of (controlled) block encoding of $\alpha_i(t) \pi H_i/2$ to construct the block encoding of $\sum_{i=1}^m \frac{\pi}{2m} \alpha_i(t)  H_i$. Hence, the complexity of obtaining such linear combination is
$$ \mathcal{O} \Big(  \sum_{i=1}^m T_W \deg(\alpha_i(t))  \big( N d_i ||H_i|| + \log \frac{1}{\epsilon} \big) \log \frac{1}{\epsilon} \Big)= \mathcal{O}\Big( m T_W \big( N d_{\max} ||H||_{\max} +  \log \frac{1}{\epsilon}   \big) \log \frac{1}{\epsilon} \Big)  $$
where $d_{\max} = \max_i \{d_i\}$ and $||H||_{\max} = \max_i \{ ||H_i|| \}$. The last step is performing exponentiation to obtain the block encoding of $\exp\Big(- \sum_{i=1}^m \frac{\pi}{2m} \alpha_i(t) H_i \Big)$, which takes further complexity $ \mathcal{O}\Big(\frac{\pi s}{2m} || \sum_{i=1}^m \alpha_i(t) H_i || \log \frac{1}{\epsilon}  \Big) = \mathcal{O}( \frac{\pi s}{2m}|| \int_0^t H(s) ds || \log \frac{1}{\epsilon} )$ where we have used $\sum_{i=1}^m \alpha_i(t) H_i = \int_0^t H(s) ds$ and $d$ is the sparsity of $\sum_{i=1}^m \alpha_i(t) H_i$, which is also the sparsity of $H(t)$. So the total complexity of obtaining $\exp\Big(- \sum_{i=1}^m \frac{\pi}{2m} \alpha_i(t) H_i \Big)$ is 
$$ \mathcal{O}\Big(  m T_W \big( N  d_{\max} ||H||_{\max} +  \log \frac{1}{\epsilon}   \big) \log \frac{1}{\epsilon} \cdot \big( \frac{\pi d}{2m}  N  \Big|\Big| \int_0^t H(s) ds\Big|\Big|  + \log \frac{1}{\epsilon} \big)  \Big) $$
To obtain $\exp\big(-i\sum_{i=1}^m \pi \alpha_i H_i\big)$, we simply note that it is equal to products of $m$ operators $\exp(- \sum_{i=1}^m \frac{\pi}{2m} \alpha_i H_i)$. Thus, the total complexity is:
$$ \mathcal{O}\Big( m  \big( d N \Big|\Big| \int_0^t H(s) ds  \Big|\Big| + m  \log \frac{1}{\epsilon} \big) T_W \big( N d_{\max} ||H||_{\max} +  \log \frac{1}{\epsilon}   \big) \log \frac{1}{\epsilon} \Big)  $$
We remind that $T_W$ is the time required to implement a single-qubit rotational gate with angle $\arccos(t)$. In reality, such gate is realized by turning (time-independent) Hamiltonian for a corresponding time $\arccos(t)$, hence the time required to execute it would be $T_W = \mathcal{O}(1)$. So the overall complexity to simulate time-dependent Hamiltonian $H(t)$ is: 
$$\mathcal{O}\Big( m   \big( d N \Big|\Big| \int_0^t H(s) ds  \Big|\Big| + m  \log \frac{1}{\epsilon} \big)  \big( N d_{\max} ||H||_{\max} +  \log \frac{1}{\epsilon}   \big) \log \frac{1}{\epsilon} \Big)  $$
Comparing with the optimal time-independent Hamiltonian simulation result \cite{low2017optimal, low2019hamiltonian, berry2015hamiltonian}, the above result is matching the optimal regime for most parameters, such as sparsity, time and norm of composing Hamiltonian. For error dependence, the framework outlined here is nearly optimal, as the scaling is polylogarithmical in inverse of error dependence $1/\epsilon$. One may immediately ask, in what context it is possible to obtain optimal scaling in $1/\epsilon$. A key step in our work is to assume the ability to simulate $\exp(-iH_i t)$ for all $H_i$, then use it to obtain the block encoding of $\pi H_i/2$. If, for example, we have access to the oracle specifying (sparse) $H_i$, then its block encoding $H_i/ (d_i ||H_i||_{\max})$ could be easily obtained, using the result in \cite{gilyen2019quantum}. Another example is the model of linear combination (LCU), where each $H_i$ is a implementable unitary. Then by definition of block encoding, such unitary block encoding itself, and we can use it to execute the framework outlined. In these scenarios, we can remove the step involving simulation of $H_i$ and ``taking logarithmic'' of such unitary. Hence, the complexity is then $\mathcal{O}\Big( m  \big(d N   \Big|\Big| \int_0^t H(s) ds  \Big|\Big| + m  \log \frac{1}{\epsilon} \big) \Big) $. 

\noindent
\textbf{Discussion and Comparison.} We remark that most attempts in simulating time-dependent system fall into either sparse-access (SM) or linear combination of unitary (LCU) model \cite{berry2014exponential, poulin2011quantum, kieferova2019simulating,low2018hamiltonian, berry2020time}. Despite these models are distant from ours, it is worth discussing existing works and point out how our method is advantageous. In SM model, the input assumption is an oracle access to entries of a $d$-sparse Hamiltonian $H(t)$ at any given time. The best algorithm for this model as far as we know is \cite{berry2020time}, Section 3.3, which achieves complexity $\mathcal{O}\Big( N  d^4 \log \frac{1}{\epsilon} \big( \int_0^t \Big|\Big| H(s)  \Big| \Big| ds \big)^2 \Big)$. An improvement to $\mathcal{O}\Big( N \big( \log \frac{1}{\epsilon} \big)  d \int_0^t \Big|\Big| H(s)  \Big| \Big| ds  \Big)$ is possible with an extra oracle that computes the so-called inverse change-of-variable, e.g., see Section 4.2 in \cite{berry2020time}. It is apparent that $\int_0^t ||H(s)|| ds  > || \int_0^t H(s) ds ||$, so our method is polynomially better than the method of Section 3.3 and slightly better than the method of Section 4.2 in \cite{berry2020time}. In the LCU model, a Hamiltonian admits the decomposition $H(t) = \sum_{l=1}^L \alpha_{l}(t) U_l$ where $U_l$ is both Hermitian and unitary with known implementation circuit. This model bears somewhat similarity to ours, with an interchange of notation $l \leftrightarrow m$, $\gamma_i(t) \leftrightarrow \alpha_l(t)$, $H_i \leftrightarrow U_l$, except that $U_l$ is unitary. Indeed, if all $\{ U_l \}_{l=1}^L$ commute, then our method can also applies to this model. The method of Section 3.3 in Ref.~\cite{berry2020time} has scaling $\mathcal{O}\big(  \frac{||\alpha||_1^2}{\epsilon}   \big)$ where $||\alpha||_1 =\sum_{i=1}^L \int_0^t ds  \alpha_l(s) ||U_l||_{\infty}$. Thus our work achieves an exponential improvement in error tolerance. Within the same LCU model, the method of Section 4.2 in Ref.~\cite{berry2020time} has scaling $\mathcal{O}\big( ||\alpha||_{\infty} L^2 \big)$, which implies that our work is quadratically better in $L$. To further extend the reach of our framework, we point out a few physical scenarios where our method can be applied. 

\noindent
\textbf{Generalized Lattice.} The first type of system that falls into our regime is generalized lattice Hamiltonian, which also appeared in \cite{childs2019nearly, haah2021quantum}. Consider a simple model where $n$ qubits are laid out on a one-dimensional lattice. The interactions between qubits are local and involve only nearest qubit. The Hamiltonian describes the system is $H = \sum_{j=1}^{n-1} H_{j,j+1} (t)$ where each $H_{j,j+1}(t) = \gamma_{j,j+1}(t) H_{j,j+1}$ with operator $H_{j,j+1}$ being time-independent, and $\gamma_{j,j+1}(t)$ is a continuous/integrable function, featuring time-dependent interaction between nearby qubits. In order to apply our framework, we need $H(t)$ to be commuted at different times, which implies that all $\{H_{j,j+1}(t) \}_{j=1}^{n-1}$ commute at different times. It is apparent that $H_{j,j+1}(t)$ and $H_{k,k+1}(t)$ commute for $|j-k| \geq 2$. Hence, it is only required that $H_{j,j+1}(t)$ commute with $H_{j+1,j+2}(t)$, which implies that $H_{j,j+1}$ commute with $H_{j+1,j+2}$ for all $j$, because for different times $t_1,t_2$, $\gamma_{j,j+1}(t_1) \gamma_{j+1,j+2}(t_2) = \gamma_{j+1,j+2}(t_2) \gamma_{j,j+1}(t_1)$. What kind of structure should admit this property? Let us give an example where the composing Hamiltonian is a tensor product of Pauli operators. For an $n$-site lattice, let $H_{j,j+1} = \Ibb_{1,2,...,j-1} \otimes \sigma^x_j \otimes \sigma^y_{j+1} \otimes \Ibb_{j+2,...,n}$ and $H_{j+1,j+2} = \Ibb_{1,2,...,j} \otimes \sigma^y_{j+1} \otimes \sigma^z_{j+2} \otimes \Ibb_{j+3,...,n}$. It is easy to see that they commute. Likewise, if $H_{j-1,j} = \Ibb_{1,2,...,j-2} \otimes \sigma^y_{j-1} \otimes \sigma^x_j \otimes \Ibb_{j+1,...,n}$ then it also commutes with $H_{j,j+1}$ and $H_{j+1,j+2}$ as defined previously. This example has illuminated that for as long as two consecutive terms $H_{j,j+1}$ and $H_{j+1,j+2}$ share a common $j+1$-th term, then all $\{ H_{j,j+1} \}_{j=1}^{n-1}$ commute, thus $H(t)$ at different times commute. In this model, each $H_{j,j+1}$ can be directly block-encoded, thus the complexity of our approach is $\mathcal{O}\Big(  n t \big( \Big|\Big| \int_0^t H(s) ds \Big|\Big| + \log \frac{1}{\epsilon}  \big)   \Big)$. The lower bound has been established in \cite{haah2021quantum} as $\Omega ( nt )$, which implies that our framework is optimal in $t$ (up to some factor) and nearly optimal in $n$ -- the number of sites. Compared to the first-order product formula in \cite{childs2019nearly,tran2020destructive} that achieves complexity $\mathcal{O}\big(  \frac{nt^2}{\epsilon} \big)$, our framework is quadratically better in time and exponentially better in error tolerance $1/\epsilon$, which is significant. This implies that our method can be efficiently realized in the near-term era. 

\noindent
\textbf{Floquet systems.} It is a quantum system with a time-dependent Hamiltonian that is periodic, i.e., $H(t+T) = H(t) $. A wide range of physical phenomena are captured in this time-periodic model, including topological phases and time crystals \cite{kitagawa2010topological, rudner2013anomalous, else2016floquet, khemani2016phase, harper2020topology}. The periodic property of $H(t)$ allows a Fourier expansion $H(t) = \sum_{m \in \mathbb{Z}} e^{-i m \omega t} H_m $ where $\omega = \frac{2\pi}{T}$. Suppose that at certain $H_m = 0$ for $|m| > |m_{\max}| = \mathcal{O}(1)$ and $H_m$ commutes with $H_k$ for $m \neq k$, which implies that $H(t)$ commute at different times. This condition is also assumed in a relevant work \cite{mizuta2023optimal}. In this setting, it is straightforward to have $\int_0^t \exp(-im \omega t) = \frac{-1}{im\omega}\exp(-im\omega t)$. This function admits a polynomial approximation, e.g., Jacobi-Anger expansion, as used in \cite{low2017optimal,low2019hamiltonian}. Thus, our framework can be executed in a straightforward manner. If $H_m$ is simulable, then the complexity for simulating the periodic system as described above is $\mathcal{O}\Big( t \log^3 \frac{1}{\epsilon}   \Big)$. Otherwise, if $H_m$ is provided directly as some unitary (similar to $H_{j,j+1}$ being products of Pauli operators in the generalized lattice case), then the complexity is further optimized as $\mathcal{O}\Big( t \log \frac{1}{\epsilon} \Big)$, which is optimal in both time and error tolerance. Compared to relevant work \cite{mizuta2023optimal}, our method achieves almost the same complexity, but is much simpler. 

\noindent
\textbf{Phase Transition.} Phase transition is a fundamental physical phenomenon, where the properties of some physical system change with the change of external conditions. For example, a quantum Ising model has Hamiltonian $H = -J \sum_{i,j} \sigma_i^z \otimes \sigma_j^z - h_x \sum_i \sigma_i^x$ exhibits ferromagnetic phase when $h_x$ is small, and transits to paramagnetic when $h_x$ exceeds a certain value. It is thus straightforward to model the Hamiltonian as $H(t) = -J \sum_{i,j} \sigma_i^z \otimes \sigma_j^z - h_x(t) \sum_i \sigma_i^x $ where $h_x(t)$ could be a linear function of $t$, indicating that the change of magnetic field is linear in time. An interesting phenomenon occurs upon sudden turning on of a magnetic field. We can model $h_x(t)$ as a rectangle function, which admits a polynomial approximation according to \cite{gilyen2019quantum}, and we quote the result here for simplicity: 
\begin{lemma}[Polynomial approximation to rectangle function]
    Let $\delta,\epsilon \in (0,1/2)$ and $t \in [-1,1]$. There is an even polynomial $P \in \mathbb{R}[x]$ of degree $\mathcal{O}( \frac{1}{\delta} \log \frac{1}{\epsilon'})$ such that $|P(x)| \leq 1$ for all $x \in [-1,1]$ and
    \begin{align}
        \begin{cases}
            P(x) \in [0,\epsilon] \text{ for all  } x \in [ -1,-t- \delta] \cup [t+\delta,1]  \\
            P(x) \in [1-\epsilon,1] \text{ for all } x \in [-t+\delta , t- \delta] 
        \end{cases}
    \end{align}
\end{lemma}
Many other interesting phase transition phenomena, for example, the Bose-Einstein condensate in optical lattice, can also occur by an abrupt change of magnetic field and thus can be modeled in the same way. This implies that our framework can be applied to the study of phase transitions. 

\noindent
\textbf{Conclusion.} In this work we have proposed a new framework for simulating time-dependent Hamiltonian. Our method is simple, using elementary techniques from quantum signal processing and its generalization, the quantum singular value transformation framework. Our method applies to Hamiltonian that commutes with itself at different times. Despite this restriction, we have pointed out a few physical settings where the commutativity condition is met. As a consequence, our newly introduced framework provides simpler solution but achieves optimal/nearly optimal complexity for simulating generalized lattice and time-periodic system, which are two important models that potentially host a diverse array of physical phenomena. In particular, our proposal has also open up a new revenue in studying phase transition on a quantum computer, which is of fundamental interest.  

\subsection*{Acknowledgement}
The author thanks Trung V. Phan for insightful discussion that motivated this project. We acknowledge support from Center for Distributed Quantum Processing, Stony Brook University.

\bibliography{ref.bib}{}
\bibliographystyle{unsrt}

\clearpage
\newpage
\onecolumngrid
\appendix

\end{document}